\documentclass[fleqn,twoside]{article}

\usepackage{espcrc2}
\usepackage{graphicx}
\usepackage[figuresright]{rotating}

\newcommand{\AmS}{{\protect\the\textfont2
  A\kern-.1667em\lower.5ex\hbox{M}\kern-.125emS}}

\title{Analysis of $N^{\star}$ spectra using matrices of correlation
       functions based on irreducible baryon operators}

\author{LHP Collaboration: S. Basak\address[UMD]{Department of Physics,
        University of Maryland, College Park, MD 20742, USA}%
        \thanks{Presented by S. Basak},
    R. Edwards\address[JLab]{Thomas Jefferson National Accelerator
    Facility, Newport News, VA 23606, USA},
    R. Fiebig\address[FIU]{Department of Physics, Florida
    International University, Miami, FL 33199, USA},
    G.T. Fleming\addressmark[JLab]\address[Yale]{Sloane Physics
    Lab, Yale University, 217 Prospect St, New Haven, CT 06520, USA},
    U.M. Heller\address[APS]{American Physical Society, One
    Research Road, Ridge, NY 11961, USA},
    C.~Morningstar\address[CMU]{Department of Physics, Carnegie
    Mellon University, Pittsburgh, PA 15213, USA},
    D. Richards\addressmark[JLab],
    I. Sato\addressmark[UMD]
    and
    S. Wallace\addressmark[UMD]}

\begin{document}

\begin{abstract}
We present results for ground and excited-state nucleon masses in
quenched lattice QCD using anisotropic lattices. Group theoretical
constructions of local and nonlocal straight-link irreducible
operators are used to obtain suitable sources and sinks. Matrices
of correlation functions are diagonalized to determine the
eigenvectors. Both chi-square fitting and Bayesian inference
with an entropic prior are used to extract masses from the
correlation functions in a given channel. We observe clear
separation of the excited state masses from the ground state mass.
States of spin $\ge 5/2$ have been isolated by use of $G_2$
operators.
\end{abstract}

\maketitle

\section{INTRODUCTION}

Reproducing the spectrum of baryon resonances with spin-1/2 and
spin-3/2 and both parities is an important test of lattice QCD.
For that we require three-quark operators that transform
irreducibly under the spinorial rotation group of the lattice
\cite{collins}. Local and nonlocal straight-link operators
corresponding to irreducible representations (IRs) $G_{1g,u}$,
$H_{g,u}$ and $G_{2g,u}$ are constructed to obtain suitable
sources and sinks \cite{ikuro}. Here we analyze $N^\star$ spectra
using these operators.

To determine the $N^\star$ excited states, a matrix of correlation
functions is computed in the quenched approximation to QCD using
irreducible baryon interpolating operators
$\overline{B}_i(\vec{x},t)$ of definite quantum numbers,
\begin{equation}
C_{ij}(t) = \sum_{\vec{x}} \langle 0 \vert T \big ( B_i(\vec{x},t)
\overline{B}_j(\vec{0},0) \big ) \vert 0 \rangle.
\label{geneig}
\end{equation}
The $C_{ij}(t)$ matrices for $G_1$ and $H$ states for each parity
are constructed using local operators, smeared local operators
and smeared straight-link operators. For $G_2$, we have one
operator which is of smeared straight-link type (Tables 1, 3 in
\cite{ikuro}).

The computation of masses of the lowest-lying resonances is
based on the variational method applied to the matrix of
correlation functions. In this paper we solve the generalized
eigenvalue equation,
\begin{equation}
C_{ij}(t) \, V_j^{(\alpha)}(t) = \lambda^{(\alpha)}(t,t_0) \,
C_{ij}(t_0) \, V_j^{(\alpha)}(t)
\label{geveq}
\end{equation}
and determine eigenvectors $V^{(\alpha)}(t)$ for each $t$, with
$t_0$ close to the source time. Then the masses of $N^\star$
states correspond to the eigenvalues of Eq.~(\ref{geveq}):
$\lambda^{(\alpha)}(t,t_0) \overrightarrow{\;\; _{t\gg t_0} }
\; e^{-m^{(\alpha)}(t-t_0)}$ \cite{luscher}. The effective
masses are determined from 
\begin{equation}
m_{{\rm eff}}^{(\alpha)} = \ln \left [ \frac{\lambda^{(\alpha)}
(t,t_0)} {\lambda^{(\alpha)}(t+1,t_0)} \right ]
\overrightarrow{\;\; _{t\gg t_0} } \;\;m^{(\alpha)}.
\label{meff}
\end{equation}

\begin{figure}[t]
\includegraphics[width=75mm,height=60mm,angle=0]{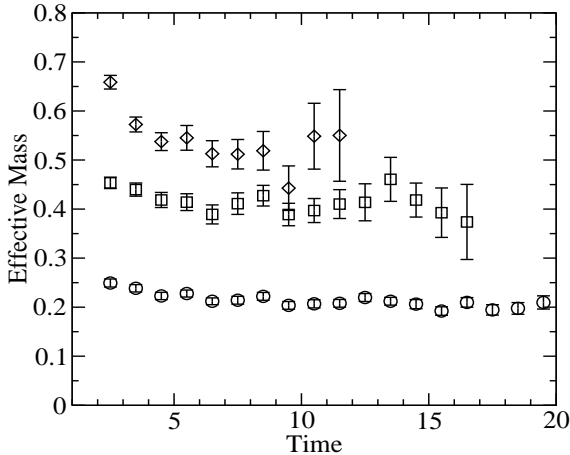}
\vspace{-12mm}
\caption{$G_{1g}$ effective masses for a selected few
low-lying states.}
\label{g1g}
\vspace{-4mm}
\end{figure}

Another way to extract spectrum information is to calculate the
spectral mass density $\rho(\omega)$ from lattice correlation
functions using the Maximum Entropy Method (MEM) \cite{fiebig},
\begin{equation}
C(t,t_0) \,\longrightarrow \, \int d\omega \,\rho(\omega)
e^{-\omega(t-t_0)}.
\end{equation}
One of the advantages of MEM is that
it utilizes data on a wide range of available time slices
and has been shown to yield results even for noisy data
\cite{fiebig}. This feature may be helpful in extracting
masses of excited states.

\section{RESULTS}

We use an ensemble of 287 quenched, anisotropic $16^3 \times 64$
lattices with renormalized anisotropy $\xi=3.0$ and $\beta=6.1$,
corresponding to $a_t^{-1} = 6.0$ GeV \cite{robert}. We use the
anisotropic Wilson action. The parameters of the Wilson fermion
action are tuned nonperturbatively so as to satisfy the continuum
dispersion relation $E({\bf p})^2=E(0)^2 + c({\bf p})^2{\bf p}^2$
at a pion mass $m_\pi \simeq 500$ MeV. To improve the coupling of
operators to the lower mass states we employ gauge-covariant
smearing of the quark fields on both source and sink:
$\tilde{\psi}(x) = (1 + \sigma^2 \tilde{\Delta} (\tilde{U}) /
4N)^N \psi(x)$, where $\tilde{\Delta}(\tilde{U})$ is the three
dimensional Laplacian and $\tilde{U}$ denotes APE-smeared $SU(3)$
link variables. The parameters used to smear the quark fields are
$\sigma=3.6$ and $N=32$.

\begin{figure}[t]
\includegraphics[width=75mm,height=60mm,angle=0]{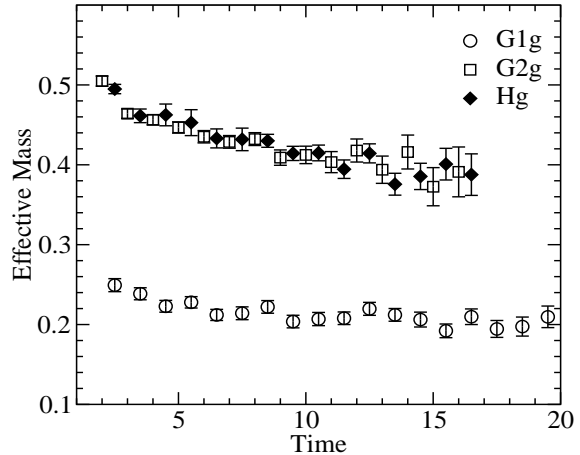}
\vspace{-12mm}
\caption{Lowest positive parity effective masses for the
IRs $G_{1g}$, $H_g$ and $G_{2g}$.}
\label{g1g2hg}
\vspace{-4mm}
\end{figure}

The effective masses are calculated from $10 \times 10$ $G_{1g}$
matrices with $t_0=2$. In Fig.~\ref{g1g} we choose to show a few
low-lying states that are clearly separated. However, the details
of the states above the ground state are under study. The plot
shows a good plateau for the ground state and statistically
significant splittings for a couple of excited states.

\begin{figure}[t]
%\vspace{-12mm}
\includegraphics[width=75mm,height=60mm,angle=0]{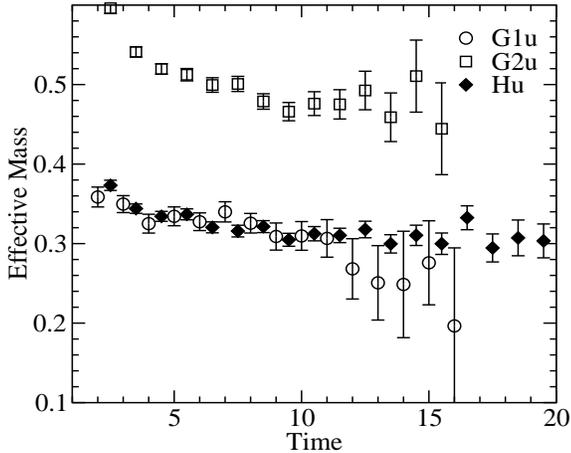}
\vspace{-12mm}
\caption{Lowest negative parity effective masses
for the IRs $G_{1u}$, $H_u$ and $G_{2u}$.}
\label{g1g2hu}
\vspace{-4mm}
\end{figure}

In Figs. \ref{g1g2hg} and \ref{g1g2hu} we have collected the
effective mass plots of the lowest states of both parities for
$G_1$, $G_2$ and $H$. The ratio of lowest masses for $G_{1g}$
and $G_{1u}$ is roughly in accordance with experiment, for 
spin-1/2 states, the $G_{1u}$ mass being higher. The effective
masses for $H_{g,u}$ are obtained using a $7 \times 7$ matrices
of correlation functions. The lowest negative parity $H_u$ state
has smaller mass than the lowest positive parity $H_g$ state.
This is compatible with the pattern found in nature for spin-3/2.
However, the masses of $H_u$ and $G_{1u}$ overlap within errors.
\begin{table}[b]
\caption{Numerical estimates from single-exponential
fit to $\lambda^{(\alpha)}(t,t_0)$ corresponding to the lowest
masses of $N^\star$ spectra.}
\label{mlist}
\vspace{2mm}
\renewcommand{\arraystretch}{1.2} % enlarge line spacing
\begin{tabular}{|l|c|c|c|}
\hline
IRs & Fit range & $m_{{\rm eff}}$ & $\sim$ Mass (MeV) \\
\hline
$G_{1g}$ & 9 -- 20 & 0.208 (4) & 1250 \\
$G_{1u}$ & 8 -- 12 & 0.321 (4) & 1930 \\
$H_{g}$ &  9 -- 12 & 0.410 (2) & 2460 \\
$H_{u}$ &  6 -- 15 & 0.315 (4) & 1890 \\
$G_{2g}$ & 9 -- 14 & 0.409 (7) & 2450 \\
$G_{2u}$ & 8 -- 15 & 0.475 (7) & 2850 \\
\hline
\end{tabular}
\end{table}
Our result for $G_2$ masses also reveals reasonable separations
of the $G_{2g}$ and $G_{2u}$ masses, $G_{2g}$ being lower. From
Fig. \ref{g1g2hg}, it is evident that the effective mass for $H_g$
(allowed spin $3/2^+$, $5/2^+$, $\cdots$) is very similar to that for
$G_{2g}$ (allowed spin $5/2^+$, $7/2^+$, $\cdots$). These states are
orthogonal. One possibility for this is that the lowest $H_g$ state
has spin-$3/2^+$ and its mass is accidentally close to that of the
lowest $G_{2g}$ state. Another possibility is that the lowest $H_g$
state is spin-$5/2^+$, in which case the same state must be present
in $H_g$ and $G_{2g}$, but not in $G_{1g}$. Study over different
values of lattice spacing is required to decide.

\begin{figure}[t]
\includegraphics[width=60mm,height=75mm,angle=90]{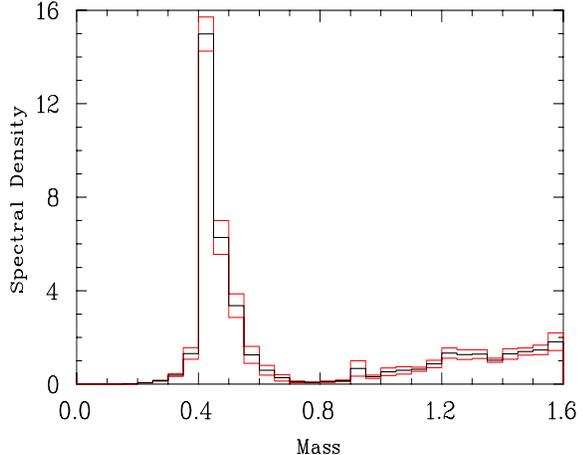}
\vspace{-12mm}
\caption{Example of a spectral density function from
a MEM analysis of $G_{2g}$ correlation functions.}
\label{mem}
\vspace{-4mm}
\end{figure}

Finally, we present the $G_{2g}$ MEM spectral function in Figure
\ref{mem}. We find that the peak of the spectral density roughly
corresponds to the effective mass value.

In Table \ref{mlist} we summarize our preliminary estimates
of the lowest masses for the different representations extracted
from single-exponential fits to the $\lambda^{(\alpha)}(t,t_0)$ of
Eqn.~\ref{geneig}. The effective masses for the lowest states of
$G_1$, $H$ and $G_2$ for both parities, whether obtained from the
variational method or preliminary MEM analysis, show a
spectrum of distinct $N^\star$ masses. However, the behavior
of the spectrum with $m_\pi$ and the sensitivity of the spectrum
to variations in the lattice volume has yet to be studied.

This work is supported by US National Science Foundation under
Awards PHY-0099450 and PHY-0300065, and by US Department of Energy
under contract DE-AC05-84ER40150 and DE-FG02-93ER-40762.


\begin{thebibliography}{9}
\bibitem{collins} C. Morningstar {\em et al.}, Nucl. Phys. B (Proc.
Suppl.) 129 (2004) 236 and these proceedings
\bibitem{ikuro} I. Sato {\em et al.} Nucl. Phys. B (Proc. Suppl.) 129
(2004) 209 and these proceedings
\bibitem{luscher} C. Michael, Nucl. Phys. B259 (1985) 58;
M. L\"{u}scher {\em et al.}, Nucl. Phys. B339 (1990) 222; C.R.
Alton {\em et al.}, Phys. Rev. D47 (1993) 5128
\bibitem{fiebig} H.R. Fiebig, Phys. Rev. D65 (2002) 094512
\bibitem{robert} R.G. Edwards {\em et al.}, Nucl. Phys. B (Proc.
Suppl.) 119 (2003) 305
\end{thebibliography}
\end{document}